




\input phyzzx

\overfullrule=0pt

\def\M{{\cal M}}
\def\p{\partial}
\def\wt{\widetilde}

\def\define#1#2\par{\def#1{\Ref#1{#2}\edef#1{\noexpand\refmark{#1}}}}
\def\con#1#2\noc{\let\?=\Ref\let\<=\refmark\let\Ref=\REFS
         \let\refmark=\undefined#1\let\Ref=\REFSCON#2
         \let\Ref=\?\let\refmark=\<\refsend}

 \let\refmark=\NPrefmark

\define\BLUM
J. Blum, preprint EFI-94-04 (hep-th/9401133).

\define\WITTENSIGMA
E. Witten, Nucl. Phys. {\bf B202} (1982) 253.

\define\ATHIT
M. Atiyah and N. Hitchin, Phys. Lett. {\bf 107A} (1985) 21; Phil. Trans.
R. Soc. Lond. {\bf A315} (1985) 459; The Geometry and Dynamics of Magnetic
Monopoles, Princeton University Press (1988).

\define\FONT
A. Font, L. Ibanez, D. Lust and F. Quevedo, Phys. Lett. {\bf B249}
(1990) 35; S.J. Rey, Phys. Rev. {\bf D43} (1991) 526.

\define\GAUNT
J. Gauntlett, Nucl. Phys. {\bf B400} (1993) 103; preprint EFI-93-09
(hep-th/9305068).

\define\GIBMAN
G. Gibbons and N. Manton, Nucl. Phys. {\bf B274} (1986) 183.

\define\HARSTR
J. Harvey and A. Strominger, Comm. Math. Phys. {\bf 151} (1993) 221.

\define\SREV
A. Sen, preprint TIFR-TH-94-03 (hep-th/9402002), and references therein.

\define\ORTIN
T. Ortin, Phys. Rev. {\bf D47} (1993) 3136.

\define\MONTOLIVE
C. Montonen and D. Olive, Phys. Lett. {\bf B72} (1977) 117;
P. Goddard, J. Nyuts and D. Olive, Nucl. Phys. {\bf B125} (1977) 1.

\define\OSBORN
H. Osborn, Phys. Lett. {\bf B83} (1979) 321.

\define\SCHSEN
J. Schwarz and A. Sen,  Nucl. Phys. {\bf B411} (1994) 35;
Phys. Lett. {\bf B312} (1993) 105.

\define\SENBOGOM
A. Sen, Mod. Phys. Lett. {\bf A8} (1993) 2023.

\define\WITTENOLIVE
E. Witten and D. Olive, Phys. Lett. {\bf B78} (1978) 97.

\define\WITTENTHETA
E. Witten, Phys. Lett. {\bf B86} (1979) 283.

\define\MANTONBOUND
N. Manton, Phys. Lett. {\bf B198} (1987) 226.

\define\GIBRUB
G. Gibbons and P. Ruback, Comm. Math. Phys. {\bf 115} (1988) 267.

\define\MANSCH
N. Manton and B. Schroers, Annals. of Phys. {\bf 225} (1993) 290.

{}~\hfill \vbox{ \hbox{hep-th/9402032}\hbox{TIFR-TH-94-08}\hbox{February,
1994}}

\title{DYON - MONOPOLE BOUND STATES, SELF-DUAL HARMONIC FORMS ON THE
MULTI-MONOPOLE MODULI SPACE, AND SL(2,Z) INVARIANCE IN STRING THEORY}

\author{Ashoke Sen\foot{e-mail addresses: sen@theory.tifr.res.in,
sen@tifrvax.bitnet}}

\address{Tata Institute of Fundamental Research, Homi Bhabha Road, Bombay
400005, India}

\abstract

Existence of SL(2,Z) duality in toroidally compactified heterotic string
theory (or in  the N=4 supersymmetric gauge theories), that includes the
strong weak coupling duality transformation, implies the existence of
certain supersymmetric bound states of monopoles and dyons. We show that
the existence of these bound states, in turn, requires the existence of
certain normalizable,
(anti-)self-dual, harmonic forms on the moduli space of
BPS multi-monopole configurations, with specific symmetry properties.
We give an explicit construction of this harmonic form on the two
monopole moduli space, thereby proving the existence of all the required
bound states in the two monopole sector.

\vfill\eject

Heterotic string theory, compactified on a six dimensional torus, has been
conjectured to possess an SL(2,Z) symmetry, part of which interchanges the
strong and weak coupling limits of the string theory\SREV. This
generalizes the strong-weak coupling duality conjecture of Montonen and
Olive\MONTOLIVE\ for the N=4 supersymmetric gauge theories\OSBORN, which
is the non-gravitational sector of the low energy effective field theory
of toroidally compactified heterotic string theory. In Ref.\SREV\ we
presented several pieces of evidence in support of this conjecture, and
also worked out several consequences of this duality symmetry. One of the
consequences is the existence of certain bound states of the known BPS
monopole and dyon states in the theory. In this paper we shall study the
criteria for the existence of these bound states. By working in a region of
the moduli space where the masses of the monopoles and dyons are small
compared to the Planck mass, we can ensure that gravity does not play an
important role in the analysis, so that our results are equally valid for
string theory, as well as for the N=4 supersymmetric gauge theories. We
shall show that the existence of these bound states requires the existence
of certain normalizable, (anti-)self-dual,
harmonic forms on the multi-monopole
moduli space with specific symmetry properties, and explicitly construct
such a form on the two monopole moduli space.

The relevant parameters characterizing the theory under consideration are
the loop expansion parameter $g^2$ and the theta angle $\theta$. In string
theory these can be related to the asymptotic values of the dilaton and
the axion fields respectively, but in N=4 supersymmetric theory they must
be treated as external parameters. It is convenient to combine the
two parameters into a single complex parameter $\lambda=(\theta/2\pi)+
ig^{-2}\equiv \lambda_1+i\lambda_2$. The electric and magnetic charges of a
state are characterized by two integers $m$ and $n$.\foot{For toroidally
compactified heterotic string theory, the unbroken gauge group at a
generic point in the compactification moduli space is U(1)$^{28}$, and we
must specify two vectors $\vec\alpha$ and $\vec\beta$ in an appropriate 28
dimensional lattice to specify the electric and magnetic charges of a
state. Since we shall be interested in the case where the vectors
$\vec\alpha$ and $\vec\beta$ are parallel to each other\SREV, we may write
$\vec\alpha=m\vec\alpha_0$, $\vec\beta=n\vec\alpha_0$, where $\vec\alpha_0$
is the smallest lattice vector along the direction of $\vec\alpha$ and
$\vec\beta$. Thus, for fixed $\vec\alpha_0$, the electric and magnetic charge
vectors are specified by the two integers $m$ and $n$. The BPS monopoles
that we shall discuss here correspond to states for which
$(\vec\alpha_0)^2=-2$ in the notation of Ref.\SREV.} Taking into account
the Witten effect\WITTENTHETA\ the physical electric
charge can be shown to be proportional to $(m +n\theta/2\pi)$
(see Ref.\SREV\ for complete expression for the electric and magnetic
charges). The
SL(2,Z) transformations are generated by matrices of the form $A\equiv
\pmatrix{p & q\cr r & s}$ with $p$, $q$, $r$, $s$ integers, satisfying
$ps-qr=1$, and act on the quantum numbers $m$ and $n$, and the coupling
constant $\lambda$\FONT, in the following way,
$$ \pmatrix{m \cr n}\to
\pmatrix {m'\cr n'} = A \pmatrix{m \cr n} =
\pmatrix{pm + qn \cr rm + sn}, \quad \quad \lambda \to {p\lambda - q\over
-r\lambda + s}. \eqn\eone
$$
The mass of a state in a sector characterized by the quantum numbers $m$
and $n$ is bounded from below by a function $f(m,n)$, known as the
Bogomol'nyi bound\WITTENOLIVE, given by\ORTIN\SENBOGOM,
$$ \Big( f(m,n)\Big)^2 = C \times \pmatrix{ m & n} \M \pmatrix{m \cr n},
\eqn\ethree
$$
where $C$ is a normalization constant,\foot{In general $C$ depends on
other modular parameters of the theory like the Higgs vacuum expectation
value, and also the vector $\vec\alpha_0$ specifying the direction of the
electric and the magnetic charge vectors. But since these parameters do
not transform under the SL(2,Z) transformation, we can treat them as
constants for the purpose of this paper.} and\SCHSEN\
$$
\M = {1\over \lambda_2} \pmatrix{ 1 & \lambda_1 \cr \lambda_1 &
|\lambda|^2}. \eqn\efour
$$
It can be easily checked that $f(m,n)$ is invariant under the SL(2,Z)
transformation \eone. Particles saturating the Bogomol'nyi bound are
annihilated by half of the 16 supersymmetry generators, and belong to a 16
component representation of the supersymmetry algebra. A class of
elementary string excitations (
{\it e.g.} the massive gauge supermultiplet of the N=4
supersymmetric theories with spontaneously broken non-abelian gauge
symmetries) are of this kind, carrying quantum numbers $(m=\pm 1, n=0)$.

Given the existence of a state with quantum number $(m=1, n=0)$, SL(2,Z)
symmetry of the theory requires the existence of other states, whose
quantum numbers are related to these by the SL(2,Z) transformation given
in Eq.\eone. The charge
quantum numbers carried by these states are given by,
$$ \pmatrix{m' \cr n'} = \pmatrix{ p & q\cr r & s}\pmatrix{1 \cr 0} =
\pmatrix{p \cr r}. \eqn \etwo $$
Note that SL(2,Z) transformation relates the charge spectrum for one value
of $\lambda$ to charge spectrum for a different value of $\lambda$, since
it acts non-trivially on $\lambda$. We shall assume that the elementary
excitations carrying charge quantum numbers $\pmatrix{1\cr 0}$ exist for
all values of $\lambda$ in the upper half plane.\foot{This follows from
the general argument given in Ref.\WITTENOLIVE\ if we assume that the
spectrum varies continuously with $\lambda$.} In that case states
carrying charge quantum numbers given in Eq.\etwo\ must also exist for all
values of $\lambda$. States with $p=0$, $r=1$ correspond to the usual BPS
monopole solution in the theory, and belong to a 16 component
supermultiplet\OSBORN. States with $p\ne 0$ and $r=1$ can be identified
with the BPS dyon states in the theory, and are also known to belong to
the 16-component supermultiplet. The states with $r\ge 2$ will be
the subject of discussion of the present paper.

Note that $p$ and $r$ cannot be arbitrary integers, but must be such that
for some integers $s$ and $q$, $ps-qr=1$. We can find integers $s$ and $q$
satisfying this requirement if and only if $p$ and $r$ are relatively
prime. Thus we need to prove the existence of a {\it single} 16 component
supermultiplet, saturating the Bogomol'nyi bound, for each pair of values
of $p$ and $r$ which are relatively prime.

The significance of the requirement of $p$ and $r$ being relatively prime
was discussed in Ref.\SREV. For completeness we shall review the main
points.
If $p$ and $r$ are not relatively prime, then there exist
integers $p_0$, $r_0$ and $n$ such that $p = n p_0$, $r=n r_0$. In this
case a state carrying quantum numbers $\pmatrix{p \cr r}$ and saturating
the Bogomol'nyi bound has the same mass as the sum of masses of $n$ states
carrying quantum numbers $\pmatrix{p_0 \cr r_0}$ and saturating the
Bogomol'nyi bound. Thus even if there exists a state saturating the
Bogomol'nyi bound and having quantum numbers $\pmatrix{p \cr r}$, it is
energetically indistinguishable from the lowest energy state in the
continuum containing $n$ particles with quantum numbers $\pmatrix{p_0 \cr
r_0}$. On the other hand, if $p$ and $r$ are relatively prime, then the
mass of any state with these quantum numbers and saturating the
Bogomol'nyi bound will be necessarily lower than the sum of the masses of
any two or more states whose charge quantum numbers add up to
$\pmatrix{p\cr r}$. This is seen by using the
triangle inequality,
$$ f(p,r) \le f(p_1, r_1) + f (p_2, r_2), \eqn\esix $$
for any $p_i$, $r_i$ satisfying $p=p_1+p_2$, $r=r_1+r_2$. The equality
holds if and only if $p_1/r_1=p_2/r_2=p/r$, which is impossible if $p$ and
$r$ are relatively prime. Thus for $p$ and $r$ relatively prime, a state
carrying quantum numbers $\pmatrix{p \cr r}$ and saturating the
Bogomol'nyi bound, has an energy strictly less than the lowest energy
state in the
continuum. Such states, if they exist, may be regarded as bound states of
$r$ monopoles and/or dyons, each carrying a single unit of magnetic charge.

In order to look for these bound states, we must study the quantization of
the collective modes of the $r$ monopole solution.  We shall carry out our
analysis for the case where $\theta=0$, but the extension to non-zero
values of $\theta$ should be straightforward following the procedure of
Ref.\WITTENTHETA.  Fortunately this problem has already been solved in the
recent work of\break
\noindent Refs.\con\HARSTR\GAUNT\BLUM\noc. The bosonic part of the
configuration space $M_r$ of the $r$ monopole solution is known to have
the structure\ATHIT\
$$
M_r = R^3 \times {S^1 \times M_r^0\over Z_r}, \eqn\eseven $$
where $R^3$
denotes the configuration space of the center of mass coordinate of the
monopole, $S^1$ is labeled by the coordinate $\chi$ conjugate to the
total charge of the monopole, and $M_r^0$ is a non-trivial $4(r-1)$
dimensional space. The group of transformations $Z_r$ is generated by an
element $g$
that acts on the coordinate of $S^1$ by a shift $\chi \to
\chi + (2\pi/r)$, and also has a non-trivial action on the manifold
$M_r^0$. In order to carry out the quantization on $M_r$, we can first
carry out the quantization on $R^3\times S^1\times M_r^0$ and then
restrict ourselves to $Z_r$ symmetric states. If we
are looking at a sector with total electric charge quantum number $p$,
then the wave-function on $R^3\times S^1$ is proportional to
$e^{ip\chi}$ and picks up a multiplicative factor of $e^{2\pi i p/r}$
under the action of $g$. Thus the component of the wave-function on
$M_r^0$ must pick up a factor of $e^{-2\pi i p/r}$ under the action of
$g$.

For the monopoles in N=4 supersymmetric theories, it was shown in
Refs.\GAUNT\BLUM\ that for each bosonic collective coordinate, there are
fermionic collective coordinates represented by a
two component Majorana fermion. Let $X^\alpha$
($1\le \alpha\le 4(r-1)$) and $Y^a$ ($1\le a\le 4$) denote the coordinates
on $M_r^0$ and $R^3\times S^1$ respectively, and let $\lambda^\alpha$ and
$\eta^a$ be the corresponding two component fermionic collective
coordinates. The metric on $R^3\times S^1$ is known to be flat. Let us
normalize the coordinates $Y^a$ such that this metric is equal to
$\delta_{ab}$. Also, let us denote by $g_{\alpha\beta}$ the metric on
$M_r^0$, by $\Gamma^\alpha_{\beta\gamma}$ the corresponding Christoffel
symbol, and by $R_{\alpha\beta\gamma\delta}$ the Riemann curvature tensor.
The dynamics of the
collective coordinates is described the lagrangian\GAUNT\BLUM\
$$ L = L_0 + L_{int}, \eqn \eeight $$
where,
$$ L_0 = {1\over 2} \p_0 Y^a \p_0 Y^a + {i\over 2} \bar\eta^a \gamma^0
\p_0 \eta^a, \eqn\enine $$
$$ L_{int} = {1\over 2} g_{\alpha\beta} \p_0 X^\alpha \p_0 X^\beta +
{i\over 2} g_{\alpha\beta} \bar \lambda^\alpha \gamma^0 D_0 \lambda^\beta
+ {1\over 12} R_{\alpha\beta\gamma\delta} \bar \lambda^\alpha
\lambda^\gamma \bar \lambda^\beta \lambda^\delta, \eqn\eten $$
and,
$$ D_0 \lambda^\beta = \p_0 \lambda^\beta + \Gamma^\beta_{\alpha\gamma}
\p_0 X^\alpha \lambda^\gamma. \eqn\eeleven $$
The quantization of $L_0$ is straightforward and has been discussed
before\GAUNT, so let us just mention the main features. If $P_a$ denote
the momenta conjugate to the coordinates $Y^a$, then the Hamiltonian is
given by ${1\over 2} P_a P_a$. The vector $(P_1, P_2, P_3)$ can be
interpreted as the spatial momenta, and $P^4$ may be interpreted as the
electric charge, up to certain normalization constants. The corresponding
eigenvalue of the Hamiltonian is known to saturate the Bogomol'nyi
bound\GAUNT\ (in the semi-classical approximation of large $g$ that we are
using). Also, since the Hamiltonian is independent of the fermionic
coordinates, there is a large degeneracy. Since each $\eta^a$ is two
component, there are eight fermionic coordinates altogether; we can
quantize them by treating half of them as creation operators and half of
them as annihilation operators. This gives a $2^4=16$ fold degeneracy of
each state.

Before we go on to quantize $L_{int}$, let us try to determine what we
should expect in order for SL(2,Z) to be a valid symmetry of the theory.
Since the contribution from $H_0$ already saturates the Bogomol'nyi bound,
we need an eigenstate of $H_{int}$ of zero eigenvalue, in order that the
combined state saturates the Bogomol'nyi bound. Furthermore, since we want
the final state to have a 16-fold degeneracy, and since the quantization
of $H_0$ already gives rise to this degeneracy, we need to have a
unique eigenstate of $H_{int}$ with zero eigenvalue for each value of the
electric charge quantum number $p$, for which $p$ and $r$ are relatively
prime. (Note that although $p$ does not appear explicitly in the
expression for $H_{int}$, it enters through the requirement that the
wave-functions on $M_r^0$ must pick up a phase $e^{-2\pi i p/r}$ under the
action of the generator $g$ of $Z_r$ transformations.)

Let us now discuss quantization of $L_{int}$. Fortunately, this
has already been carried out by
Witten\WITTENSIGMA. It was shown that the states in the Hilbert space of
this system are in one to one correspondence to the differential forms on
$M_r^0$, and a zero energy eigenstate corresponds to a harmonic form on
$M_r^0$. Thus, in order to establish the SL(2,Z) invariance of the
spectrum, we need to establish the existence of a {\it unique}
harmonic form  on $M_r^0$, which picks up a phase of $e^{-2\pi i p/r}$
under the action of the generator $g$ of the $Z_r$ group. Since by our
previous argument it is guaranteed to represent a true bound state, such
an harmonic form will automatically be normalizable.

Now, given a harmonic $p$ form $\omega$ on $M_r^0$ satisfying the required
properties, we can always construct a harmonic $4(r-1)-p$ form on $M_r^0$ by
taking the Hodge dual of $\omega$.
This would violate the condition that $\omega$ must be the unique harmonic
form on $M_r^0$.
The only exception is the case where
$\omega$ is an (anti-)self-dual $2(r-1)$ form. Thus we see that SL(2,Z)
invariance of the
spectrum not only demands the existence of a harmonic form with specific
$Z_r$ transformation properties,
but it also requires this harmonic form to be an (anti-)self-dual $2(r-1)$
form.

To summarize, SL(2,Z) invariance of the spectrum requires that

\noindent {\it For every integer $p$ for which $p$ and $r$ are relatively
prime,
the space $M_r^0$  must contain a normalizable,
(anti-)self-dual, harmonic $2(r-1)$
form, which picks up a phase $e^{-2\pi i p/r}$ under the action of the
generator $g$ of the $Z_r$ transformations.}

In the rest of the paper we shall discuss the construction of such a
harmonic form in the special case $r=2$. The metric on the space $M_2^0$
is known explicitly\ATHIT\GIBMAN. This space is labeled by the coordinates
$(\rho,
\theta, \phi, \psi)$  ($0\le \theta \le \pi$,
$0\le \phi\le 2\pi$, $0\le \psi \le 2\pi$) with the identification\GIBMAN,
$$ (\rho, \theta, \phi, \psi) \equiv (\rho, \pi -\theta, \pi+\phi, -\psi).
\eqn\eione $$
$\rho$ is a radial coordinate.
The metric is given by\ATHIT\GIBMAN\
$$ ds^2 = f^2 d\rho^2 + a^2 (\sigma_1)^2 + b^2 (\sigma_2)^2 + c^2
(\sigma_3)^2, \eqn\emetric $$
where $f$, $a$, $b$ and $c$ are known functions of $\rho$, and,
$$\eqalign{
\sigma_1 =& -\sin\psi d\theta + \cos \psi \sin\theta d\phi, \cr
\sigma_2 =& \cos\psi d\theta + \sin\psi \sin\theta d\phi, \cr
\sigma_3 =& d\psi +\cos\theta d\phi. }
\eqn\esigmas
$$
The $\sigma_i$'s satisfy the relation,
$$ d\sigma_i = {1\over 2}
\epsilon^{ijk} \sigma_j \wedge \sigma_k. \eqn\edsigma $$

The generator $g$ of $Z_2$ transformation acts on the coordinates $(\rho,
\theta, \phi, \psi)$ as,
$$ (\rho, \theta, \phi, \psi) \to (\rho, \theta, \phi, \psi+\pi). \eqn
\eitwo $$
Since $r=2$, the relevant values of $p$ are the set of all odd integers.
For each of these values of $p$, the action of $g$ is required to send the
(anti-)self-dual form to its negative.

Thus what we are looking for is an (anti-)self-dual harmonic two form
$\omega$, which is invariant under the transformation \eione, and goes
to its negative under the transformation \eitwo. We take the following
trial solution
$$ \omega = F(\rho) \Big(d \sigma_1 - {fa\over bc} d\rho\wedge
\sigma_1\Big). \eqn\etrial $$
By construction, this form is anti-self-dual, and satisfies all the
symmetry requirements. Thus we only need to make sure that $d\omega=0$;
since $\omega$ is anti-self-dual, this will automatically give $d^\dagger
\omega=0$. This gives,
$$ {dF\over d\rho} = -{fa\over bc} F, \eqn\eforf $$
and hence,
$$ F(\rho) = F_0 \exp (-\int_\pi^\rho {fa\over bc} d\rho'), \eqn\effinal $$
where $F_0\equiv F(\pi)$.
With the parametrization of Ref.\GIBMAN, we have the following asymptotic
form for the functions $f$, $a$, $b$ and $c$ as $\rho\to \infty$:
$$ f\simeq -1, \quad \quad a\simeq \rho, \quad \quad b\simeq \rho, \quad
\quad c\simeq -2. \eqn\easympone $$
Thus, as $\rho\to \infty$, ${fa\over bc}\to {1\over 2}$. This gives
$F\simeq F_0 \exp(-\rho/2)$ asymptotically, showing that the harmonic form
is normalizable.
This is expected, since, according to our previous argument, the energy of
this state is strictly less than that of the continuum.

Finally, we need to ensure that the harmonic form $\omega$ is non-singular
near the `Bolt', $\rho \simeq \pi$. The correct choice of coordinates in
this region is $\wt\rho\equiv \rho -\pi$, and a new set of Euler angles
$\wt\theta$, $\wt\phi$ and $\wt\psi$\GIBMAN. The one forms $\sigma_i$ are
given in this coordinate system by,
$$\eqalign{
\sigma_1=& d\wt\psi +\cos\wt\theta d\wt\phi, \cr
\sigma_2 =& -\sin\wt\psi d\wt\theta + \cos\wt\psi \sin\wt\theta d\wt\phi,
\cr
\sigma_3 =& \cos\wt\psi d\wt\theta + \sin\wt\psi \sin\wt\theta d\wt\phi.}
\eqn\eboltone
$$
The functions $f$, $a$, $b$ and $c$ are approximated by,
$$ f\simeq -1, \quad \quad a\simeq 2\wt\rho, \quad \quad b\simeq \pi, \quad
\quad c\simeq -\pi. \eqn\easympbolt $$
Thus from Eq.\emetric\ we see that, for $\rho\simeq \pi$,
$$ ds^2 = d\wt\rho^2 + 4\wt\rho^2 (d\wt\psi + \cos\wt\theta d\wt\phi)^2 +
\pi^2 (d\wt\theta^2 + \sin^2 \wt\theta d\wt\phi^2). \eqn\emetricbolt$$

Let us now express $\omega$ given in \etrial\ in this coordinate system
near $\rho=\pi$. From Eq.\effinal\ we see that $F\to F_0$ as $\rho\to \pi$,
so that $\omega$ takes the form
$$ \omega = F_0 \Big(\sin\wt\theta d\wt\phi\wedge d\wt\theta - {2\over
\pi^2} \wt\rho d\wt\rho\wedge (d\wt\psi +\cos\wt\theta d\wt\phi)\Big).
\eqn\omegabolt
$$
{}From the form of the metric \emetricbolt\ we see that the one forms
$\sin\wt\theta d\wt\phi$, $d\wt\theta$, $d\wt\rho$, and $\wt\rho (d\wt\psi
+ \cos\wt\theta d\wt\phi)$ are all well defined near the Bolt. This, in
turn, shows that the two form $\omega$ is well defined near the Bolt.

Our result establishes the existence of supersymmetric bound states in the
two monopole sector, with sixteen fold degeneracy,
for each odd value of the total charge of the
system.\foot{Note that since the anti-self-dual harmonic form $\omega$
cannot be regarded as a (2,0) or a (0,2) form, it does not correspond to a
supersymmetric bound state in the N=2 supersymmetric theories\GAUNT.}
This is in prefect agreement with the predictions of
SL(2,Z) symmetry. Generalizing these results to the multi-monopole case
will require proving the statement made in the paragraph above Eq.\eione.

Note added: The self-dual 2-form $\omega$ has been used by Gibbons and
Ruback\GIBRUB\ for different purpose earlier, and has been further
explored in Ref.\MANSCH.
Bound states of monopoles
and dyons in non-supersymmetric theories have been discussed by
Manton\MANTONBOUND.

\refout

\end